\documentclass[aps,prl,twocolumn,showpacs,floatfix]{revtex4}
\usepackage{graphicx}
\usepackage{dcolumn}
\usepackage{amsthm,amsmath}

\begin{document}
\title{Quadrupole Shifts for the $^{171}$Yb$^+$ Ion Clocks: Experiments versus Theories}
\author{D. K. Nandy and B. K. Sahoo }
\affiliation{Theoretical Physics Division, Physical Research Laboratory, Ahmedabad-380009, India}
\affiliation{Indian Institute of Technology Gandhinagar, Ahmedabad, India}
\email{dillip@prl.res.in}
\email{bijaya@prl.res.in}
\date{Received date; Accepted date}
 
\begin{abstract}
Quadrupole shifts for three prominent clock transitions, $[4f^{14} 6s] ^2S_{1/2} \rightarrow [4f^{14} 5d] ^2D_{3/2}$,   
 $[4f^{14} 6s] ^2S_{1/2} \rightarrow [4f^{14} 5d] ^2D_{5/2}$ and $[4f^{14} 6s] ^2S_{1/2} \rightarrow [4f^{13} 6s^2] ^2F_{7/2}$, 
in the Yb$^+$ ion are investigated by calculating the quadrupole moments ($\Theta$s) of the $5d_{3/2,5/2}$ and $4f_{7/2}$ states
using the relativistic coupled-cluster (RCC) methods. We find an order difference in the $\Theta$ value of the $4f_{7/2}$ state 
between our calculation and the experimental result, but our result concur with the other calculations that are carried 
out using different many-body methods than ours. However, our $\Theta$ value of the $5d_{3/2}$ state is in good agreement 
with the available experimental result and becomes more precise till date to estimate the the quadrupole shift of the $[4f^{14} 6s] 
^2S_{1/2} \rightarrow [4f^{14} 5d] ^2D_{3/2}$ clock transition more accurately. To justify the accuracies in our calculations, we 
evaluate the hyperfine structure constants of the $6s_{1/2}$, $5d_{3/2,5/2}$ and $4f_{7/2,5/2}$ states of $^{171}$Yb$^+$ ion using the 
same RCC methods and compare the results with their experimental values. We also determine the lifetime of the $5d_{3/2}$ state to 
eradicate the scepticism on the earlier measured value as claimed by a recent experiment.
\end{abstract}

\maketitle

\begin{figure}[t]
\includegraphics[width=6.5cm, height=5.1cm, clip=true]{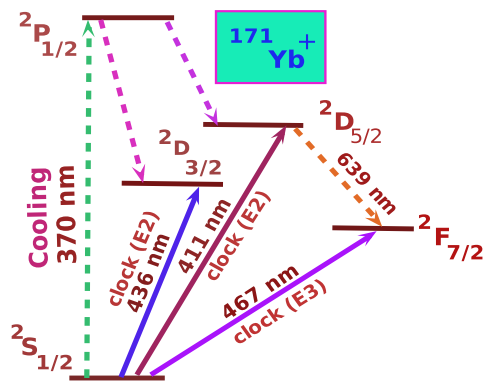}
\caption{(color online) Schematic view of the energy levels and the clock transitions in the $^{171}$Yb$^+$ ion.}
\label{fig1}
\end{figure}

A single trapped Al$^+$ ion is the most accurate atomic clock till date \cite{chou1} implying that one of the singly charged 
ions is capable of becoming the primary frequency standard in future provided its stability can be further improved. The other 
successful optical single ion clocks are Hg$^+$ \cite{oskay}, Ca$^+$ \cite{matsubara}, Sr$^+$ \cite{margolis}, 
Yb$^+$\cite{tamm, roberts} etc. In Yb$^+$, two quadrupole (E2) $[4f^{14}6s] ^2S_{1/2} \rightarrow  [4f^{14}5d] ^2D_{3/2}$ and 
$[4f^{14}6s] ^2S_{1/2} \rightarrow  [4f^{14}5d] ^2D_{5/2}$ transitions having optical wavelengths 436 nm and 411 nm, respectively, and 
an octupole (E3) $[4f^{14}6s] ^2S_{1/2} \rightarrow [4f^{13}6s^2] ^2F_{7/2}$ transition with optical wavelength 467 nm are considered 
for the clock measurements, see Fig. \ref{fig1}, in many laboratories around the globe \cite{tamm, roberts, imai, 
subha}. Since the field-induced frequency shifts in the $[4f^{13}6s^2] ^2F_{7/2}$ state is very low and it is also highly meta-stable 
\cite{huntemann}, it makes the above octupole transition as an instinctive choice to think as the most precise and stable next genre optical 
clock. Although the lifetime of the $[4f^{13}6s^2] ^2F_{7/2}$ state is very long ($>$ 6 years) which cannot be considered as the interrogation 
time during the clock frequency measurement, instead its probe interaction time ($\sim$ 10 s) serves this purpose \cite{huntemann}. On 
the otherhand, the lifetimes of the metastable $[4f^{14}5d] ^2D_{3/2}$ and $[4f^{14}5d] ^2D_{5/2}$ states are about 55 $ms$ and 7 $ms$, 
respectively \cite{yu, schacht} and can be used as the interrogation times in the clock transitions involving these states. Owing to 
these facts, many other important studies like parity nonconservation \cite{bijaya1, rahaman}, quantum information \cite{olmschenk}, 
variation of the fine structure constant \cite{godun} etc. using the above transitions in Yb$^+$ are also in progress.

 One of the major resources that contribute to the uncertainty budget of a clock frequency measurement is the quadrupole shift resulting  
from the stray electric field gradient $({\bf \nabla E^{(2)}})$ during the experiment \cite{itano1}. This shift can be accurately 
estimated with the precise knowledge of the quadrupole moments $(\Theta$s) of the states involved in a clock transition. This urges for 
determination of $\Theta$s for the $[4f^{14}5d] ^2D_{3/2}$,  $[4f^{14}5d] ^2D_{5/2}$ and $[4f^{13}6s^2] ^2F_{7/2}$ states ($\Theta$ is 
zero for the $[4f^{14}6s] ^2S_{1/2}$ state) of Yb$^+$ as accurately as possible. In an experiment, $\Theta$ is measured by altering 
static direct current (dc) voltage and is very difficult to obtain very precisely. The rationale to carry out the theoretical studies 
of this property are: (i) when the experimental results are not available, the calculated values can be helpful to estimate the 
quadrupole shifts, (ii) it can prevent performing auxiliary measurements for the atomic clock experiments which are very expensive and 
(iii) comparison between the measurement and a calculation serves as a tool to test the potential of the employed many-body method. 
Thus, calculations of $\Theta$s in Yb$^+$ seem to be indispensable. The previous calculations for $\Theta$s in Yb$^+$ are reported as 2.174 $ea_0^2$ 
\cite{itano} and 2.157 $ea_0^2$ \cite{latha} against the measured value 2.08(11) $ea_0^2$ \cite{schneider} for the $[4f^{14}5d] 
^2D_{3/2}$ state and for the $[4f^{13}6s^2] ^2F_{7/2}$ state the calculated values are $-0.22 \ ea_0^2$ \cite{blythe} and $-0.20 \ 
ea_0^2$ \cite{porsev} compared to the measured value $-$0.041(5)$ea_0^2$ \cite{huntemann}. Latha et al. \cite{latha} had employed the 
relativistic coupled-cluster (RCC) method while Itano \cite{itano} had used a multi-configuration Dirac-Fock (MCDF) method to calculate
these quantities. For the $[4f^{13}6s^2] ^2F_{7/2}$ state, Blythe et al. \cite{blythe} had employed the MCDF method, while Porsev et al.
\cite{porsev} report their result inconclusively using a CI method and predicting the final value as $\sim -0.1 \ ea_0^2$. In this Letter, we 
intend to perform calculations of $\Theta$s of these states including their fine structure partners $[4f^{14}5d] ^2D_{5/2}$ and 
$[4f^{13}6s^2] ^2F_{5/2}$ states by considering all possible configurations within the singles and doubles approximation in our recently 
developed \cite{yashpal,nandy} RCC (CCSD) methods. These methods are suppose to be more accurate than the truncated CI or MCDF methods 
on the physical grounds \cite{szabo, bartlett}, hence we may possibly apprehend the role of the electron 
correlations better in the determination of $\Theta$s and to elucidate plausible reasons for the discrepancies between the theoretical and 
experimental results. In addition, we calculate the magnetic dipole hyperfine constants ($A_{hf}$s) of the above states 
of $^{171}$Yb$^+$ and compare them against their experimental values to gain insights into the accuracies of our calculations. 
Furthermore, we determine the lifetime of the $[4f^{14}5d] ^2D_{3/2}$ state to eradicate the conflict about its correct value which is given differently by two separate
measurements \cite{yu, schacht}.

\begin{table*}[t]
\caption{Contributions from the CCSD methods (after dividing by the corresponding normalization factors) and comparison between the other available results of the quadrupole moments ($\Theta$s)
in $ea_0^2$ and the magnetic dipole hyperfine structure constants ($A_{hf}$s) in MHz of the low-lying states relevant to the clock 
transitions in $^{171}$Yb$^+$. Error bars are given within the parentheses.}
\begin{ruledtabular}
\begin{tabular}{lccccccccccc}
RCC &\multicolumn{2}{c}{$4f^{13}6s^2 \ ^2F_{7/2} $} 
& \multicolumn{2}{c}{$4f^{13}6s^2 \ ^2F_{5/2} $}&\multicolumn{1}{c}{$4f^{14}6s \ ^2S_{1/2} $} 
&\multicolumn{2}{c}{$4f^{14}5d \ ^2D_{3/2} $}&\multicolumn{2}{c}{$4f^{14}5d \ ^2D_{5/2} $}\\ 
term &  \\
\cline{2-3} \cline{4-5}\cline{7-8}\cline{9-10}\\  
 &$\Theta$& $A_{hf}$ &$\Theta$& $A_{hf}$ &  $A_{hf}$ &$\Theta$ & $A_{hf}$ &$\Theta$ & $A_{hf}$ \\
\hline \\
DF                                          &-0.2593 &867.66  &-0.2097 &1634.09 &7225.45  &2.440  &283.04 &3.613  &108.08  \\
\vspace{0.3mm}
$\overline{O}$-DF                           &-0.0344 &7.533   &-0.0255 &8.941   &2490.30  &-0.005 &1.95   &-0.008 &1.10    \\
\vspace{0.3mm}
$\overline{O}\Omega_{1}$                    &0.0     &0.0     &0.0     &0.0     &427.91   &-0.369 &64.30  &-0.550 &24.75  \\
\vspace{0.3mm}
$\overline{O}\Omega_{2}$                    &0.0923  &25.23   &0.0715  &87.47   &2334.97  &-0.021 &15.87  &-0.026 &-207.64 \\
\vspace{0.3mm}
$\Omega_{1}^{\dagger}\overline{O}\Omega_{1}$&0.0     &0.0     & 0.0    &0.0     &4.90     &0.046  &4.61   &0.055  &1.54    \\
\vspace{0.3mm}
$\Omega_{1}^{\dagger}\overline{O}\Omega_{2}$&0.0     &0.0     &0.0     &0.0     &-9.89    &0.0003 &3.70   &-0.0002&-13.61  \\
\vspace{0.3mm}
$\Omega_{2}^{\dagger}\overline{O}\Omega_{2}$&-0.0142 &104.17  &-0.0134 &183.78  &235.36   &-0.023 &27.60  &0.032  &16.78   \\
\vspace{0.3mm}
Final                                       &-0.216(20)&1004(100)&-0.177(50)&1914(166)&12709(400)&2.068(12)&401(14)&3.116(15) & -69(6)  \\
\hline \\
\vspace{0.3mm}
Others      &-0.22$^a$  &  &     &  & 13091$^b$      &2.174$^c$ &489$^b$   &3.244$^c$   &-96$^b$   \\  
            &-0.20$^b$                 &  &     &  &                &2.157$^d$ &400.48$^c$&            &   -12.58$^c$          \\ 
Expt. &-0.041(5)$^f$          &905.0(5)$^g$& &  &12645$^h$                 &2.08(11)$^i$&430(43)$^j$&  & -63.6(5)$^k$  

\end{tabular}
\end{ruledtabular}
\label{tab1}
$^a$\cite{blythe}, $^b$\cite{porsev}, $^c$\cite{itano}, $^d$\cite{latha}, $^e$\cite{bijaya1}, 
$^f$\cite{huntemann}, $^g$\cite{taylor}, $^h$\cite{martensson}, $^i$\cite{schneider}, $^j$\cite{engelke}, $^k$\cite{roberts2}  
\end{table*}
Theoretically quadrupole moment of a hyperfine state, $|(\gamma IJ) F M_F \rangle$, with the angular momentum $F$ and azimuthal 
component $M_F$ for the nuclear spin $I$, atomic angular momentum $J$ and $\gamma$ representing other additional information 
of the state is given by $\Theta(\gamma F)=\langle (\gamma I J) FF |\Theta^{(2)}_0| (\gamma I J) FF \rangle$ with $\Theta^{(2)}_0=
-\frac{e}{2}\sum_j(3z^2_j-r^2_j)$, the zeroth component of the quadrupole moment spherical tensor ${\bf \Theta^{(2)}}$ \cite{angel},
for which we can express \cite{itano1}
\begin{eqnarray}
\langle (\gamma IJ) FM_F |\Theta^{(2)}_q| (\gamma IJ) FM_F \rangle &=& (-1)^{F-M_F} \nonumber \\
\times \left ( \begin{matrix} F & 2 &  F \cr M_F & q & -M_F \cr 
\end{matrix}
\right ) & \times & \langle F|| {\bf \Theta^{(2)}} ||F\rangle, 
\end{eqnarray}
where $\langle F||{\bf \Theta^{(2)}}||F\rangle$ is the reduced matrix element and in the $IJ$-coupling approximation it is given by
\begin{eqnarray}
\langle F||{\bf \Theta^{(2)}} ||F\rangle &=& (-1)^{I+J+F} (2F+1) \left \{ \begin{matrix} J & 2 &  J \cr F & I & F \cr \end{matrix} 
\right \} \nonumber \\ && \times \left ( \begin{matrix} J & 2 &  J \cr J & 0 & -J \cr \end{matrix} \right )^{-1} \Theta(\gamma J)
\end{eqnarray}
for $\Theta(\gamma J)=\langle JJ |\Theta^{(2)}_0| JJ \rangle$ the quadrupole moment of the atomic state. The quadrupole shift 
in the $|(\gamma IJ) F M_F \rangle$ state due to the interaction Hamiltonian $H_Q= {\bf \nabla E^{(2)} \cdot \Theta^{(2)}}$ 
is given by \cite{itano1, brown}
\begin{eqnarray}
h \delta \nu_Q &=& \frac{-2[3M_F^2- F(F+1)]A \langle F|| {\bf \Theta^{(2)}} ||F\rangle} {[(2F+3)(2F+2)2F(2F-1)]^{1/2}} \nonumber \\
    & \times & [(3 \text{cos}^2 \beta -1) - \epsilon \ \text{sin}^2  \beta (\text{cos}^2 \alpha - \text{sin}^2 \alpha) ], \ \ \ \
\end{eqnarray}
where $\alpha$ and $\beta$ are the Euler angles used to convert the principal-axis frame to the laboratory frame,  
$\epsilon$ is known as the asymmetry parameter and $A$ is the strength of the field gradient of the applied dc voltage.

Also, the $A_{hf}$ of the  $|(\gamma IJ) F M_F \rangle$ state is given by \cite{charles}
\begin{eqnarray}
A_{hf}=\mu_N g_I\frac{\langle J||T_e^{(1)}||J\rangle}{\sqrt{J(J+1)(2J+1)}} 
\end{eqnarray}
where $g_I$ and $\mu_N$ are the gyromagnetic ratio and magnetic moment of the atomic nucleus and $T_e^{(1)}$ is the even parity tensor 
of rank one representing the electronic component of the hyperfine interaction Hamiltonian.

The lifetime of the $[4f^{14} 5d] ^2D_{3/2}$ state ($\tau_{5d3/2}$) of Yb$^+$ can be determined as
\begin{eqnarray}
\tau_{5d3/2} = \frac{1}{A_{5d3/2 \rightarrow 6s}^{M1} + A_{5d3/2 \rightarrow 6s}^{E2}},
\end{eqnarray}
where $A_{5d3/2 \rightarrow 6s}^{M1}$ and $A_{5d3/2 \rightarrow 6s}^{E2}$ are the transition probabilities from the 
$[4f^{14} 5d] ^2D_{3/2}$ state to the ground $[4f^{14} 6s] ^2S_{1/2}$ state due to the magnetic dipole (M1) and electric
quadrupole (E2) transitions, respectively. 

We consider the Dirac-Coulomb (DC) Hamiltonian to calculate the atomic wave functions which is given in the atomic unit (au) by
\begin{eqnarray}
H&=&\sum_i [ c\mbox{\boldmath$\alpha$}_D\cdot \textbf{p}_i+(\beta_D -1)c^2+
V_n(r_i) + \sum_{j \ge i} \frac{1}{r_{ij}} ], \ \ \
\end{eqnarray}
where $\mbox{\boldmath$\alpha$}_D$ and $\beta_D$ are the Dirac matrices, $c$ is the velocity of light and $V_n(r)$ is the 
nuclear potential. The considered $[4f^{14} 6s] ^2S_{1/2}$, $[4f^{14} 5d] ^2D_{3/2,5/2}$ and $[4f^{13} 6s^2] ^2F_{7/2,5/2}$ states
have the open-shell configurations, describing them using a common reference state in the the Fock-space formalism of 
the RCC theory is strenuous. For this reason, we construct two reference states, $|\Phi_0^{N-1} \rangle$ and $|\Phi_0^{N+1} \rangle$, 
using the Dirac-Fock (DF) method for the configurations $[4f^{14}]$ and $[4f^{14} 6s^2]$, respectively, with $N(=69)$ as the total 
number of electrons to calculate the above states. Here the $[4f^{14} 6s] ^2S_{1/2}$ and $[4f^{14} 5d] ^2D_{3/2,5/2}$ states can be 
determined using $|\Phi_0^{N-1} \rangle$ by attaching the respective valence electron $v$ (denoted by $|\Psi_v \rangle$) and again the 
$[4f^{14} 6s] ^2S_{1/2}$ state and the $[4f^{13} 6s^2] ^2F_{7/2,5/2}$ states can be evaluated from $|\Phi_0^{N+1} \rangle$ by 
annihilating the respective extra electron $a$ (denoted by $|\Psi_a \rangle$). The point to be noted here is that the $[4f^{14} 6s] 
^2S_{1/2}$ state obtained from $|\Phi_0^{N-1} \rangle$ and from $|\Phi_0^{N+1} \rangle$ see different DF potentials. Consequently, the 
difference in the results of this state when calculated using $|\Psi_v \rangle$ and $|\Psi_a \rangle$ at the same level of 
approximations may be able to entail the effect of the $6s$ electron in the construction of the occupied orbitals.
 
 In the Fock-space RCC formalism, only brief discussions are given here from the detailed descriptions of Refs. 
\cite{yashpal, nandy, sahoo}, we express
\begin{eqnarray}
|\Psi_v \rangle =e^{T^{N-1}+S_v} a_v^{\dagger} |\Phi_0^{N-1} \rangle=e^{T^{N-1}} \{ 1+S_v \} |\Phi_v \rangle
\end{eqnarray}
and 
\begin{eqnarray}
|\Psi_a \rangle = e^{T^{N+1}+R_a} a_a |\Phi_0^{N+1} \rangle = e^{T^{N+1}} \{ 1+R_a \} |\Phi_a \rangle,
\end{eqnarray}
where $T^{N-1}$ and $T^{N+1}$ excite the core electrons from the new reference states $|\Phi_v \rangle$ and $|\Phi_a \rangle$, respectively, to account 
for the electron correlation effects and the $S_v$ operator annihilates the valence electron $v$ that was appended by $a_v^{\dagger}$ 
and creates a virtual orbital along with carrying out excitations of the core electrons from $|\Phi_0^{N-1} \rangle$ while the $R_a$ 
operator regenerates the core electron $a$ by annihilating another core electron elsewhere along with creating excitations of other core electrons from $|\Phi_0^{N+1} \rangle$. As was mentioned before, the core orbitals of $|\Phi_0^{N-1} \rangle$
do not see the interaction with the valence electron $v$. This effect along with the core-valence correlations are accounted through the 
contraction of $T^{N-1}$ and $\{1+S_v\}a_v^{\dagger}$. Analogously, the core electrons of $|\Phi_0^{N+1} \rangle$ see an extra effect 
from the spin pairing partner of $a$ which are removed through the product of $T^{N+1}$ and $\{1+R_a\}a_a$. Obviously, the core orbitals 
of $|\Phi_0^{N+1} \rangle$ are more relaxed here. The singles and doubles excitations in the CCSD methods are denoted by defining 
$T^{L}=T_1^{L} + T_2^L$ with $L=N-1$ and $L=N+1$ for the attachment and detachment cases, respectively, $S_v=S_{1v} + S_{2v}$ and 
$R_a=R_{1a}+R_{2a}$. Contributions from the important triples are estimated perturbatively \cite{nandy, sahoo} by contracting the DC 
Hamiltonian with $T_2^{N-1}$ and $S_{2v}$ in the electron attachment procedure and with $T_2^{N+1}$ and $R_{2a}$ in the detachment 
approach to account as the uncertainties due to the neglected triples.

The matrix element of a physical operator $O$ between the $|\Psi_f \rangle$ and $|\Psi_i \rangle$ states (or the expectation value with
$|\Psi_f \rangle=|\Psi_i \rangle$) are determined in our RCC method by 
\begin{eqnarray}
\frac{\langle \Psi_f | O | \Psi_i \rangle}{\sqrt{\langle \Psi_f|\Psi_f\rangle \langle \Psi_i|\Psi_i\rangle}} 
&=& \frac{\langle \Phi_f | \{ 1+ \Omega_f^{\dagger}\} \overline{O} \{ 1+ \Omega_i\} |\Phi_i\rangle}{ \sqrt{ {\cal{N}}_f 
{\cal{N}}_i  }}, \ \ \
\label{eqn25}
\end{eqnarray}
where $\overline{O}=e^{T^{L \dagger}} O e^{T^L}$ and ${\cal N}_i= (1+ \Omega_i^{\dagger}) \overline{\cal{N}} (1+ \Omega_i)$ with 
$\overline{\cal N} = e^{T^{L \dagger}} e^{T^L}$ and $\Omega_i$ is either $S_i$ for $L=N-1$ or $R_i$ for $L=N+1$. Evaluation 
procedures of these expressions are described elsewhere \cite{nandy, sahoo}.

\begin{figure}[t]
\includegraphics[width=8.8cm, height=8.5cm, clip=true]{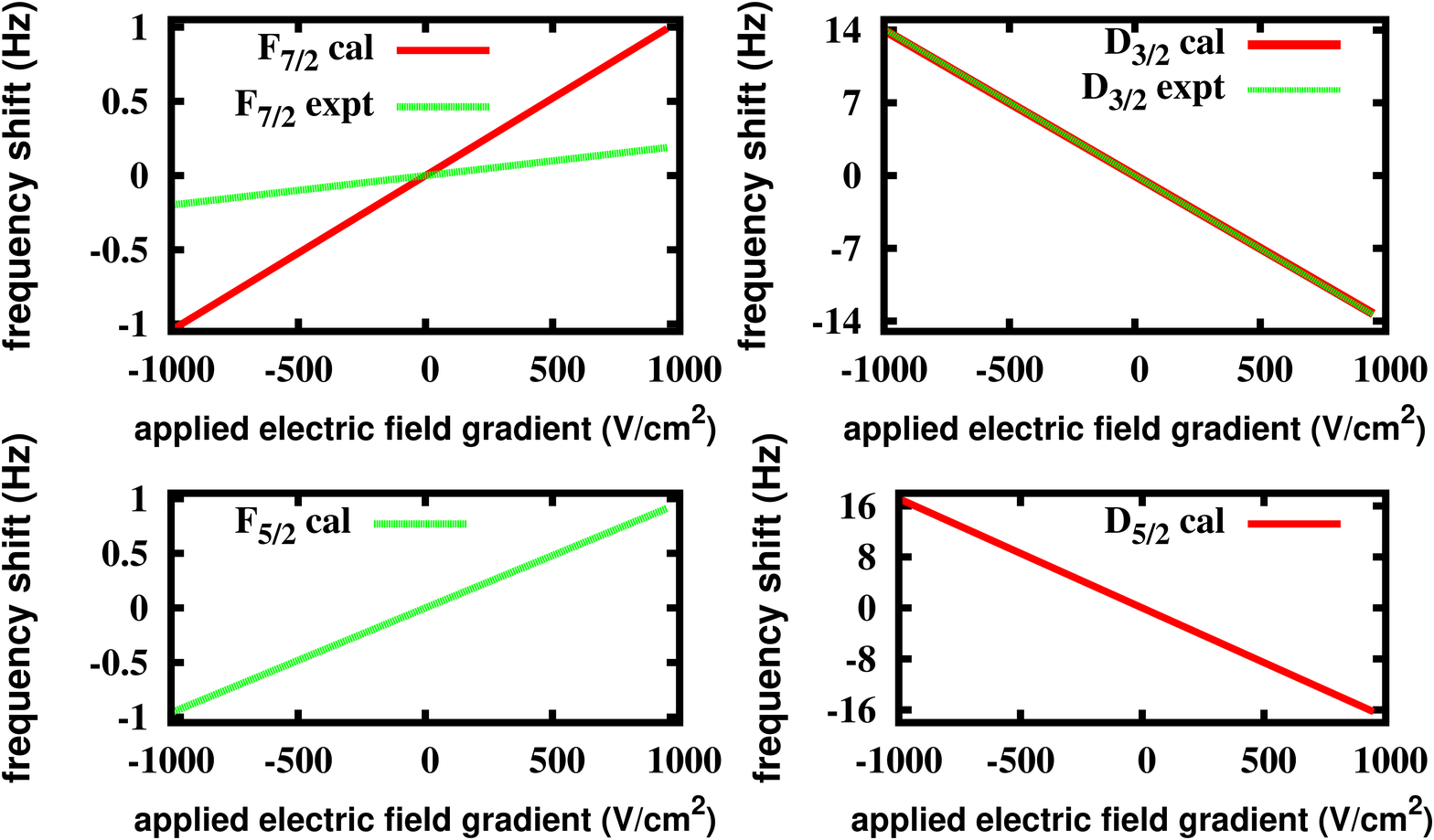}
\caption{(color online) Quadrupole frequency shifts of the $[4f^{14}5d] ^2D_{3/2}(F=2)$, $[4f^{14}5d] ^2D_{5/2}(F=2)$, $[4f^{13}6s^2] 
^2F_{7/2}(F=3)$ and $ [4f^{13}6s^2] ^2F_{5/2}(F=3)$ hyperfine states for $M_F=0$ with respect to the $[4f^{14}6s] ^2S_{1/2}(F=0)$ state 
against the electric field gradient $A$ using the calculated and measured $\Theta$ values.}
\label{fig2}
\end{figure}
In Table \ref{tab1}, we present $\Theta(\gamma J)$ values for all the considered states of $^{171}$Yb$^+$ from our calculations and 
others along with the $A_{hf}$ results and compare them with the available measurements. We also give contributions from the DF method 
and from the individual CCSD (including complex conjugate (c.c.)) terms along with the estimated upper-bounds to the uncertainties
within the parentheses in the same table. As seen, our final $\Theta$ values are almost in agreement with the other calculations and 
experimental results and also more precise, except for the $\Theta$ value of the $[4f^{13}6s^2] ^2F_{7/2}$ state. Although the 
calculations of Ref. \cite{latha} are carried out using the similar method as ours, but in the present work we have used a 
self-consistent procedure to account for the contributions from the non-truncative $\overline{O}$ series in contrast to Ref. 
\cite{latha}, in which the terms are terminated at finite number of $T^{N-1}$ operators.  Our $A_{hf}$ results seem to be agreeing with 
the experimental values within their reported error bars, which are determined using $g_I=0.98734$ \cite{stone}. The result for the 
$[4f^{14}6s] ^2S_{1/2}$ state is given only from the electron detachment method in the table. We obtain this result as 13234(900) MHz 
using the attachment method, which along with the $A_{hf}$s of the $[4f^{14}5d] ^2D_{3/2,5/2}$ states are improved over our previously 
reported results \cite{bijaya1} due to consideration of the above self-consistent procedure and larger basis set. We find 
the difference between the results of the $[4f^{14}6s] ^2S_{1/2}$ state from the two approaches, that we have considered, are very 
significant and the detachment theory gives more accurate result. Agreement between our $A_{hf}$ result of the $[4f^{13}6s^2] 
^2F_{7/2}$ state with its experimental value implies that this method is able to provide the wave functions with sufficient accuracy 
indicating that its $\Theta$ value is of similar accuracy. Therefore, the large differences between the theoretical and experimental 
results of the $[4f^{13}6s^2] ^2F_{7/2}$ state $\Theta$ values are not understandable evidently. Our intuitive guess is that 
this discrepancy could emanate, plausibly, from some unpredictable contributions arising through the triples or other higher
excitations although such signatures were obscured in our study. Its value for the hyperfine $[4f^{13}6s^2] ^2F_{7/2}(F=3)$ state, in 
which the actual measurement has been performed, yields as $-0.19(2)$ $ea_0^2$. This value is almost same with the atomic state $\Theta$
value and again far away from $-0.041(5)$ $ea_0^2$ to possibly presume that it corresponds to the hyperfine state. This, therefore, calls for
another experimental verification and more rigorous theoretical studies including higher level excitations to expunge the above 
ambiguity. Moreover, we also give the $\Theta$ of the fine structure partner, $[4f^{13}6s^2] ^2F_{5/2}$, of the above state so that its
value can be independently probed by other methods in order to cross-check our calculations. Considering our calculated $\Theta$ values for 
all the states, we plot in Fig. \ref{fig2} the quadrupole frequency shifts ($\delta \nu$) of the $[4f^{14}5d] ^2D_{3/2}(F=2)$, 
$[4f^{14}5d] ^2D_{5/2}(F=2)$, $[4f^{13}6s^2] ^2F_{7/2}(F=3)$ and $ [4f^{13}6s^2] ^2F_{5/2}(F=3)$ hyperfine states for $M_F=0$ with 
respect to the $[4f^{14}6s] ^2S_{1/2}(F=0)$ state against different $A$ values and compare them with the results estimated using the 
available experimental $\Theta$ values. These results can be used to reduce the uncertainties in the clock transitions of Yb$^+$ and 
for the further experimental investigations. 

We also obtain M1 and E2 line strengths as $2.5\times 10^{-7}$ au and $110.25$ au, respectively, for the $[4f^{14}5d] ^2D_{3/2} 
\rightarrow [4f^{14}6s] ^2S_{1/2}$ transition from our calculations. Combining these values with the experimental energies, it yields 
$A_{5d3/2 \rightarrow 6s}^{M1}=2.07\times 10^{-5}$ $s^{-1}$ and $A_{5d3/2 \rightarrow 6s}^{E2}=19.70$ $s^{-1}$. Using these
results, we get $\tau_{5d3/2}=50.78(50)$ $ms$ which is in very good agreement with the experimental result $52.7(2.4)$ $ms$ of 
Ref. \cite{yu} and repudiate the argument by the latest experiment, which observes $\tau_{5d3/2}=61.8(6.4)$ $ms$ \cite{schacht}, 
about underestimate of the systematics in the former measurement \cite{yu}.

We acknowledge PRL 3TFlop HPC cluster for carrying out the computations.

\end{document}